\documentclass[10pt]{iopart}
\usepackage{rawfonts}

\makeatletter	   
\renewcommand{\ps@plain}{%
     \renewcommand{\@oddhead}{
\textrm{Partially Directed Paths in a Wedge}\hfil\textrm{\thepage}}%
     \renewcommand{\@evenhead}{\@oddhead}%
     \renewcommand{\@oddfoot}{}
     \renewcommand{\@evenfoot}{\@oddfoot}}
\makeatother     

\input prepictex
\input pictex
\input postpictex

\newcommand{\Ref}[1]{(\ref{#1})}

\newcommand{\LL}{{I\hskip -1.10mm L}}

\newcommand{\mod}{\,\hbox{mod}\,}

\begin{document}

\title[Directed Paths in a Wedge]{Directed Paths in a Wedge}

\author{E J Janse van Rensburg$^\dagger$\footnote[3]{To
whom correspondence should be addressed (rensburg@yorku.ca)},
T Prellberg$^\ddagger$,
A Rechnitzer$^\star$,}

\address{$\dagger$ 
Department of Mathematics and Statistics,
York University, 4700 Keele Street, Toronto, 
Ontario, M3J 1P3, Canada (rensburg@yorku.ca) \\
$\ddagger$ School of Mathematical Sciences, Queen Mary,
University of London,  Mile End Road, London E1 4NS, UK 
(t.prellberg@qmul.ac.uk) \\
$\star$  Department of Mathematics,
The University of British Columbia,
Vancouver, B.C. V6T 1Z2 
(andrewr@math.ubc.ca)
}

\begin{abstract}
Directed paths have been used extensively in the 
scientific literature as a model of a linear polymer.  
Such paths models in particular the conformational entropy of 
a linear polymer and the effects it has on the free
energy.  These directed models are simplified versions
of the self-avoiding walk, but they do nevertheless give
insight into the phase behaviour of a polymer, and also
serve as a tool to study the effects of conformational
degrees of freedom in the behaviour of a linear polymer.
In this paper we examine a directed path model of a 
linear polymer in a confining geometry (a wedge).  The
main focus of our attention is $c_n$, the number of
directed lattice paths of length $n$ steps which takes
steps in the North-East and South-East directions and 
which is confined to the wedge $Y=\pm X/p$, where
$p$ is an integer.  In this paper we examine the case
$p=2$ in detail, and we determine the generating function 
using the iterated kernel method.  We also examine the
asymtotics of $c_n$. In particular, we show that
$$
c_n = [0.67874\ldots]\times 2^{n-1}(1+(-1)^n)  
 + O\left((4/3^{3/4})^{n+o(n)}\right) 
 + o\left((4/3^{3/4})^n\right) 
$$
where we can determine the constant $0.67874\ldots$ to
arbitrary accuracy with little effort.
\end{abstract}


\submitto{JPA}

\maketitle

\section{Introduction}
\label{introduction}
Lattice paths and lattice walks have long been used as
models of conformational entropy in linear polymers
\cite{F55,F69}.  Perhaps the most famous of these
models is the self-avoiding walk \cite{MS93,JvR00}.
The most fundamental quantity in this model is $w_n$,
the number of self-avoiding walks of length $n$ steps
from the origin in the hypercubic lattice.  Is it known
that $w_n = \mu^{n+o(n)}$, where $\mu$ is the growth
constant of the self-avoiding walk, and that the limit
$\lim_{n\to\infty} w_n^{1/n} = \mu$ exists
\cite{HM54}.  The self-avoiding walk is a non-Markovian
model, and it is generally very difficult to extract its
properties by rigorous or even by numerical means.

Directed paths, in particular models in two dimensions,
are generally recurrent models which can often be solved
exactly by a renewal type argument.  A particular
example is the Dyck path which renews itself each time
its visits the line $Y=X$ in the square lattice.  This
property introduces a translational invariance in the
model, which can be used directly to solve for the
generating function as a root of a quadratic polynomial,
see for example reference \cite{F68}.  This general
observation holds for other models, including a model
of directed paths above the line $Y=2X$ \cite{G86}
and partially directed lattice paths including 
bargraph paths \cite{BGW91,BGW92,PB95,PO95,PFF88} and 
other related models such as Motzkin paths \cite{DV84}.

\begin{figure}[h]
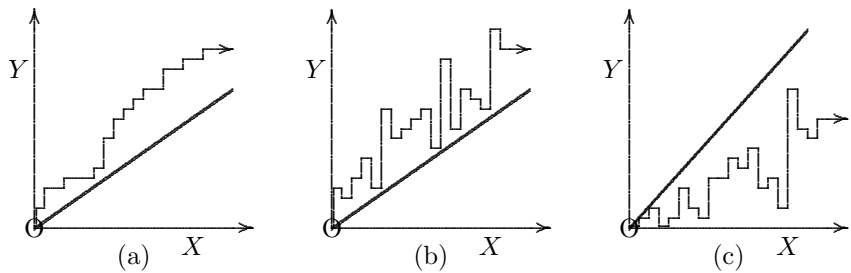

\input fig1
\caption{(a) A directed path in the wedge formed by the
$Y$-axis and the line $Y=rX$. (b) A partially directed path 
in the wedge formed by the $Y$-axis and the line $Y=rX$.}
\label{fig1}
\end{figure}

Confining a lattice path to a wedge in the square lattice
introduces complexities which oftens makes the model
harder to solve.  These models are directed versions
of the self-avoiding walk confined to a wedge \cite{HW85}, which
in turn is a model of a linear polymer in a confined geometry.
A Dyck path is perhaps the simplest directed path model of
a path confined in a wedge. In figure \ref{fig1}(a)
a more generic directed path in a wedge formed by the $Y$-axis
and the line $Y=rX$ is illustrated (where $r\geq 0$).  
This model can be solved exactly for $r\in\{0,1,2,3\}$, 
but no explicit solutions are known for other values of $r$
although the radius of convergence of the generating
function is known explicity for all $r\geq 0$,
see for example references 
\cite{DM47,D00,JvR99,JvR05,JvR05a,JvR06}.  Generally these models
pose challenging combinatorial questions.  If the line
$Y=rX$ has rational slope, then the generating function
is a root of polynomial (it is algebraic), 
and a recurrance can be determined
by a renewal type argument: The path renews itself each time
its visits the line $Y=rX$.  More generally $r$ is irrational,
and in these models there is no translational invariance
of the model along the line $Y=rX$ and a recurrence relation
for the number of paths seems out of the question.

The model in figure \ref{fig1}(a) can be generalised by 
considering a partially directed path above the line $Y=rX$
as illustrated in figure \ref{fig1}(b). 
A recurrence relation for the generating
function of these model has been written down for some
values of $r$ \cite{JvR05b}, but generally these models pose 
a significant mathematical problem.  A third variant of 
these models is illustrated in figure \ref{fig1}(c).  In this 
case a partially directed path is included in the wedge formed
by the $X$-axis and the line $Y=rX$. A model of this
type was proposed and solved in the case that $r=1$
in reference \cite{JvRRP07}.

\begin{figure}[t]
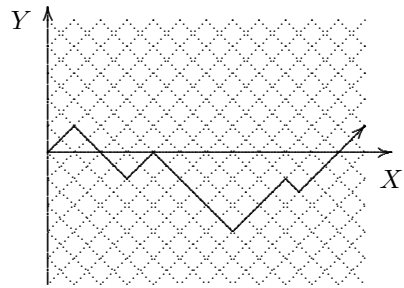

\input fig2
\caption{A fully directed path in the square lattice, starting
at the origin, and giving steps in the North-East and
South-East directions.}
\label{fig2}
\end{figure}

\begin{figure}[h]
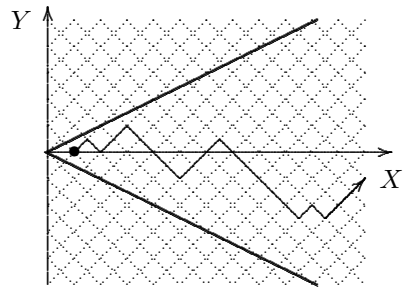

\input fig3
\caption{A directed path from the vertex $(2,0)$ in a 
symmetric wedge formed by the lines $Y= \pm X/2$.}
\label{fig3}
\end{figure}

In this paper we generalise models of fully directed paths
from the origin to a model of lattice paths in a wedge.
Consider the directed path in figure \ref{fig2} in the square
lattice which takes steps only in the North-East and
South-East directions.  The most fundamental quantity in this model is $c_n$, the number of paths from the origin of
length $n$ steps.  Obviously, in this model $c_n = 2^n$.
This path can be put into a wedge generally as 
illustrated in figure \ref{fig3}.  This is a 
symmetric wedge (about the $X$-axis), and is an
alternative and more challenging model compared to the
cases illustrated in figure \ref{fig1}(a).

We are primarily interested in a model of
directed paths starting from the vertex $(2,0)$ and confined
to the wedge formed by the lines $Y=\pm X/2$.  We determine
recurrences for the generating function using
the kernel method \cite{PBM00}, and in particular a variant
of this method developed in reference \cite{JvRRP07}
and which we will call the {\it iterated kernel method}.

The most fundamental quantity in our model is $c_n$,
the number of paths of length $n$ steps and confined to the
wedge formed by $Y=\pm X/2$ and starting from the vertex with
coordinates $(2,0)$.  We will solve for the generating function
of $c_n$, and our solution will be an alternating series
of compositions of a root of a quartic polynomial.  This
will allow us to determine $c_n$ to high accuracy: in particular,
we show that asymptotically,
\begin{equation}
\hspace{-2cm}
c_n = [0.67874\ldots]\times 2^{n-1}(1+(-1)^n)  
 + O\left((4/3^{3/4})^{n+o(n)}\right) 
 + o\left((4/3^{3/4})^n\right) 
\label{eqnasympt}
\end{equation}
where the constant $0.67874$ can be determined to hundreds of
significant digits with minimal computational effort.

In section 2 we define our models and determine a recurrence
for the generating function.  In addition, we note that
$\lim_{n\to\infty} c_n^{1/n} = 2$, we solve for the
generating function using the kernel method, and 
examine the properties of the roots of the kernel.
These results allows us to compute the constant in 
equation \Ref{eqnasympt} in section 3 by examining the
singularities in the generating function. We conclude
the paper with some final comments in section 4.

\section{Directed Paths in a Wedge}

Let $\LL$ be the square lattice of points with integer
coordinates in the plane. A directed path in this lattice 
is a path that takes only North-East (NE) and South-East 
(SE) steps.  If the path consists of $n$ steps 
(or {\it edges}), then there are $2^n$ such paths.
One such path is illustrated in figure \ref{fig2}.
Let $XY$ be the usual Cartesian coordinate system in figure
\ref{fig2}, with origin at the first vertex of the path.  Then
the edges in the directed path have length $\sqrt{2}$ each.

The directed path in figure \ref{fig2} is unconstrained by the
boundaries of the wedge $Y=\pm X$.  This model becomes more
interesting if the path is constrained by a narrower wedge
$Y = \pm X / p$, where $p\geq 1$ is an integer.  

Define the $1/p$-wedge $V_p$ by
\begin{equation}
V_p = \{ (x,y) \in \LL \, | \, \hbox{where $-x/p 
\leq y \leq x/p$} \}.
\end{equation}
Then $V_p$ is the subset of $\LL$ in the first and fourth
quadrants bounded by the lines $Y = \pm X/p$. In figure
\ref{fig3} a directed path confined to the wedge $V_p$
is denoted where $p = 2$.  This path has its first vertex at 
the point with coordinates $(2,0)$. Generally, directed 
paths in the wedge $Y = \pm X/p$ will have their first vertices 
at $(p,0)$.

\subsection{Directed paths in the Wedge formed by $Y = \pm X /2$}

In this section we determine a recurrance for the generating
function $G_2$ of directed paths giving $NE$ and $SE$ steps
from the vertex $(2,0)$ in the wedge $V_2$. Proceed by
introducing the edge generating variable $t$ and define
\begin{equation}
G_2 (a,b) = \sum_{n,u,v\geq0} c_n^{(2)} (u,v) u^a v^b t^n
\end{equation}
where $c_n^{(2)} (u,v)$ is the number of directed paths
in the wedge $V_2$ from the vertex $(2,0)$ of length $n$
and with final vertex a vertical distance $u$ from 
$Y = \lceil X/2\rceil$ and a vertical distance $v$ from 
$Y = -\lceil X/2\rceil$.

Observe that the unit of vertical distance is determined
by the fact that the squares in figure \ref{fig3} have diagonal
length $2$. For example, a NE-step will increase the $Y$-coordinate
of the endpoint by $1$, and the distance between the endpoint
and $Y=\lceil X/2 \rceil$ may either decrease by $1$ if the
initial $X$-coordinate is even, or remain unchanged otherwise.

\begin{figure}[b]
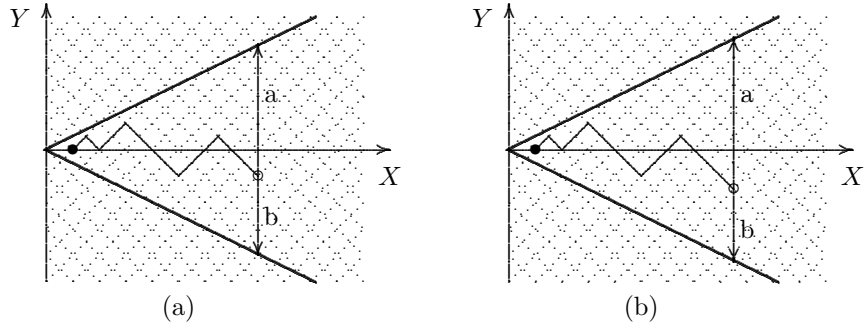

\input fig4
\caption{a 
symmetric wedge formed by the lines $Y= \pm X/2$.}
\label{fig4}
\end{figure}

Consider the examples in figure \ref{fig4}.  The vertical distances
are measured to the step-functions $Y= \pm \lceil X/2\rceil$, and
in these examples the vertical distances are
$u = 11$ and $v = 7$ (figure \ref{fig4}(a)) and
$u = 13$ and $v = 7$ (figure \ref{fig4}(b)). In addition,
$u+v$ is always even.  

We are fundamentally interested in
$c_n^{(2)} = \sum_{u,v\geq0} c_n^{(2)}(u,v)$, the number 
of paths of length $n$ in the wedge $V_2$.  The generating
variables $a$ and $b$ are introduced to account
for the vertical distance of the endpoint from the
wedge boundaries and to enable us to determine $G_2(1,1)$.

The examples of paths in figure \ref{fig4} illustrates that
the enumeration of these paths are fundamentally affected
by a parity effect.  Paths of even length (figure \ref{fig4}(a))
are either the single vertex at the point $X=2$ generated by 
$ab$, or can be generated by adding a $NE$-step or $SE$-step 
to a path of odd length.  Paths of odd length can only be 
generated by extending a path of even length by the addition of a
$NE$-step or a $SE$-step.

Let $g_0 (a,b)$ generate paths of even length, and suppose
that $g_1 (a,b)$ generates paths of odd length. Then
\begin{equation}
G_2(a,b) = g_0 (a,b) + g_1 (a,b) .
\end{equation}
Thus, by determining $g_0(a,b)$ and $g_1(a,b)$, $G_2 (a,b)$ can 
be determined.

\begin{figure}[t]
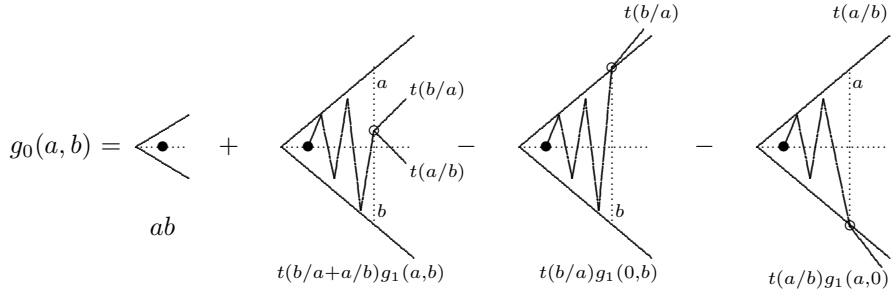

\input fig5
\caption{Determining $g_0(a,b)$.  Each path generated by
$g_0 (a,b)$ is either the single vertex at $(2,0)$
generated by $ab$, or it is generated by appending a
$NE$-edge onto a path of odd length (this gives the term
$(tab)(b/a)g_1(a,b)$), or it is generated by
appending a $SE$-edge onto a path of odd length (giving the
term $(tab)(a/b)g_1(a,b)$). Lastly, paths which step outside
the wedge must be subtracted: $(tab)(b/a)g_1(0,b)$ if the path
steps over the line $Y=X/2$ and $(tab)(a/b)g_1(a,0)$ if the
path steps over the line $Y=-X/2$.}
\label{fig5}
\end{figure}

\begin{figure}[b]
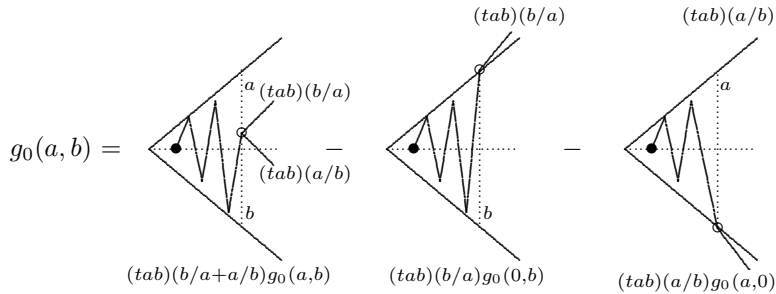

\input fig6
\caption{Determining $g_1(a,b)$.  Each path generated by
$g_1 (a,b)$ is obtained by appending a $NE$-edge or $SE$-edge
onto a path of even length.  In this case, the vertical 
distance to the wedge does not increase by a full step,
and thus no new factors of $a$ or $b$ are included, 
apart from accounting for stepping closer to the top
boundary if a $NE$-edge is added, or stepping closer to the
bottom boundary if a $SE$-edge is added.  Thus it generates
the term $t(a/b + b/a)g_0 (a,b)$ as indicated above.
Lastly, paths which step outside the wedge must be 
subtracted: $t(b/a)g_0(0,b)$ if the path
steps over the line $Y=X/2$ and $t(a/b)g_0(a,0)$ if the
path steps over the line $Y=-X/2$.}
\label{fig6}
\end{figure}

Recurrences for $g_0 (a,b)$ and $g_1(a,b)$ can be obtained by
arguing as in figures \ref{fig5} and \ref{fig6}.  The basic 
idea is to create paths generated by $g_1(a,b)$ by appending
an edge to paths of even length generated by $g_0 (a,b)$, and
to create paths generated by $g_0 (a,b)$ by appending an edge
to paths generated by $g_1 (a,b)$.  The resulting recurrances
are
\begin{eqnarray}
\hspace{-1.5cm}
g_0 (a,b) & & = ab + t(a/b+b/a)g_1(a,b) 
- t(a/b)g_1(a,0) - t(b/a) g_1 (0,b),  \nonumber \\
\hspace{-1.5cm}
g_1 (a,b) & & = t(a^2+b^2)g_0(a,b)
- ta^2g_0(a,0) - tb^2g_0 (0,b). 
\label{eqnrecur2}
\end{eqnarray}
These recurrences can be iterated to enumerate the directed
paths.  The numbers for even length paths are given in table 
\ref{table1}.

\begin{table}
\begin{center}
\begin{tabular}{||c|c||c|c||}
\hline
$n$ & $c_n$ & $n$ & $c_n$ \\
\hline
0 & 1  & 42 & 2985401474160 \\
2 & 4  & 44 & 11941093593120 \\
4 & 12 & 46 & 47764374372480 \\
6 & 48 & 48 & 191053247884320 \\
8 & 180 & 50 & 764212991537280 \\
10 & 720 & 52 & 3056816328436200 \\
12 & 2820 & 54 & 12227265313744800 \\
14 & 11280 & 56 & 48908759609676540 \\
16 & 44760 & 58 & 195635038438706160 \\
18 & 179040 & 60 & 782537580134560920 \\
20 & 713760 & 62 & 3130150320538243680 \\
22 & 2855040 & 64 & 12520579171583415840 \\
24 & 11403060 & 66 & 50082316686333663360 \\ 
26 & 45612240 & 68 & 200329075631136029040 \\
28 & 182321460 & 70 & 801316302524544116160 \\
30 & 729285840 & 72 & 3205263549296411867340 \\
32 & 2916160800 & 74 & 12821054197185647469360 \\
34 & 11664643200 & 76 & 51284202287042290859820 \\
36 & 46650808680 & 78 & 205136809148169163439280 \\
38 & 186603234720 & 80 & 820547109423871153955280 \\
40 & 746350368540 & 82 & 3282188437695484615821120 \\
\hline

\end{tabular}
\label{table1}
\end{center}
\caption{The number of directed paths in $V_2$.}
\end{table}

The number of digits in the $c_n$ in table \ref{table1} increases
linearly with $n$, which implies exponential growth of $c_n$.
It is in fact possible to prove explicitly that $c_n$ increases
proportional to $2^n$, using the techniques in reference 
\cite{HW85}. In addition, since $c_n c_m \leq c_{n+m}$ 
(concatenate a path of length $m$ with a path of length $n$ by
translating the first until its first vertex coincide with
the last vertex of the second to see this), $c_n$ is
a supermultiplicative function of $n$, and thus the limit
\begin{equation}
\lim_{n\to\infty} c_n ^{1/n} = 2
\end{equation}
exists \cite{H48}.  Thus, there is a function of $n$,
$C_0 = e^{o(n)}$, such that
\begin{equation}
c_n = C_0 2^n + o(2^n).
\end{equation}
Examination of the data in table \ref{table1} shows that
$c_n/2^n$ approaches a constant.  Assuming that $C_0$
is a constant, one may estimate it numerically.  
Dividing $c_n$ by $2^n$ and increasing
$n$ up to $n=82$ gives $C_0$ to five digits, namely
\begin{equation}
C_0 = 0.67874\ldots
\label{eqn7} 
\end{equation}

We proceed next by computing $C_0$ by solving the recurrances
in equations \Ref{eqnrecur2} above, and by examining the
singularities in $g_0 (a,b)$.

\subsection{Solving equations \Ref{eqnrecur2}}

In this section we use the iterated kernel method (see reference
\cite{JvRRP07}) to find an expression for the generating
function $g_0 (a,b)$. Simplify the recurrences by 
introducing $K = t(a/b+b/a)$ and $L=t(a^2+b^2)$. Observe 
also that $g_0(a,0) = g_0(0,a)$, and that $g_1(a,0) = g_1(0,a)$.  

Since paths of odd length generated by $g_1 (a,b)$ cannot 
intersect the boundaries of the wedge, they always will be 
weighted by factors of $a$ and $b$.  Hence, one expects 
$g_1 (a,0) = g_1 (0,b) = 0$ identically.  Ignoring this last 
observation for the moment gives the recurrances in 
slightly simplified form:
\begin{eqnarray}
g_0 (a,b) & = & ab + K\,g_1(a,b) - t(a/b)g_1(a,0) 
- t(b/a)g_1(b,0), \\
g_1 (a,b) & = &  L\,g_0(a,b) - ta^2g_0(a,0) - tb^2g_0(b,0).
\end{eqnarray}
Substitute these equations into one another, and write them
in kernel form. This gives
\begin{eqnarray}
& & \hspace{-2cm} (1-KL)g_0(a,b) \nonumber \\
& & \hspace{-1cm} = ab - ta^2K\,g_0(a,0) - tb^2K\,g_0(b,0)
- t(a/b)g_1(a,0) - t(b/a)g_1(b,0), \label{eqnkernel1} \\
& & \hspace{-2cm} (1-KL)g_1(a,b)   \nonumber \\
& & \hspace{-1cm} = Lab - ta^2g_0(a,0) - tb^2g_0(b,0)
- t(a/b)L\,g_1(a,0) - t(b/a)L\,g_1(b,0) . \label{eqnkernel2}
\end{eqnarray}
We identify the {\it kernel} $(1-KL)$ in these equations.
Generally, we say that recurrence equations 
are in {\it kernelized form} if the generating 
function and its coefficients have been collected
on the left hand side, while all other terms and boundary terms are on
the right hand side.  The kernel $(1-KL)$ may be simplified, and
then the quartic polynomial 
\begin{equation}
t^2(a^2+b^2)^2 - ab
\label{eqnquartic} 
\end{equation}
in $a$ and $b$ appears as a factor.

To proceed, consider this to be a quartic in $b$ with $a$ and
$t$ two parameters.  To solve the original recurrances, the 
roots of this quartic must be determined.  Closer examination 
shows that the four roots have the following properties:  The 
first is $\beta_0 (a)$, which is a power-series in $t$:
\begin{equation}
\beta_0 (a) = a^3 t^2 + 2a^7t^6 + 9a^{11}t^{10} + 
52a^{15}t^{14} + \ldots
\label{eqnbeta0} 
\end{equation}
The second real root is $\beta_1 (a)$, which is singular at $t=0$:
\begin{equation}
\beta_1 (a) = \frac{a^{1/3}}{t^{2/3}} - \frac{2a^{5/3}t^{2/3}}{3}
- \frac{28a^{13/3}t^{10/3}}{81} + \ldots
\end{equation}
while the two remaining roots are a complex conjugate pair:
\begin{equation}
\beta_\pm (a) = - \frac{a^{1/3}(1\mp i \sqrt{3})}{2t^{2/3}}
+ \frac{a^{5/3}t^{2/3} (1\pm i \sqrt{3})}{3} - \frac{a^3t^2}{3}
+ \ldots
\end{equation}

Substituting $b = \beta_0 (a) \equiv \beta_0$ in the 
kernelized equations in \Ref{eqnkernel1} and \Ref{eqnkernel2}
produces two equations for $g_0 (a,0)$,
$g_0 (b,0)$, $g_1 (a,0)$ and $g_1 (b,0)$:
\begin{eqnarray}
\hspace{-2cm}
K\left(ta^2g_0(a,0) + t\beta_0^2g_0(\beta_0,0)\right)
+\left(t(a/\beta_0) g_1(a,0) + t(\beta_0/a) g_1 (\beta_0,0)\right)
 & & = a\beta_0 , \\
\hspace{-2cm}
\left(ta^2g_0(a,0) + t\beta_0^2g_0(b,0)\right)
+L\left(t(a/\beta_0) g_1(a,0) + t(\beta_0/a) g_1 (\beta_0,0)\right)
 & & = La\beta_0 .
\end{eqnarray}
The solution of this linear system is
\begin{eqnarray}
\hspace{1cm}
ta^2 g_0(a,0) + t\beta_0^2 g_0 (\beta_0,0) & & = La\beta_0 , \\
t(a/\beta_0)g_1(a,0) + t(\beta_0/a)g_1(\beta_0,0) & & = 0 .
\end{eqnarray}
From the second of these it follows that
\begin{equation}
g_1(\beta_0(a),0) = - \frac{a^2}{\beta_0^2}\, g_1 (a,0)
\end{equation}
and since this generating function cannot be negative, the
conclusion is that
\begin{equation}
g_1 (a,0) = 0
\end{equation}
identically, as claimed above.  The first solution above gives
\begin{equation}
g_0 (a,0) = \frac{L(a,\beta_0)\beta_0}{ta} - \frac{\beta_0^2}{a^2}\,
g_0 (\beta_0,0).
\end{equation}

Proceed by defining $\beta^{(n)} (a) = \left( \beta_0 \circ 
\beta_0 \circ \ldots \circ \beta_0 \right) (a)$ to be the
composition of $\beta_0$ $n$-times with itself.  Define
$\beta^{(0)} (a) = a$, then the last equation may be written as
\begin{equation}
g_0 (\beta^{(n-1)},0) = \frac{L(\beta_0^{(n-1)},\beta_0^{(n)})
\beta_0^{(n)}}{t\beta_0^{(n-1)}}
- \left(\frac{\beta_0^{(n)}}{\beta^{(n-1)}}\right)^2
g_0 (\beta^{(n)},0).
\end{equation}
This may be iterated to obtain a solution for $g_0 (a,0)
= g_0(\beta_0^{(0)},0)$:
\begin{equation}
g_0 (a,0) = \frac{1}{ta^2}
 \sum_{n=0}^\infty (-1)^n L(\beta_0^{(n)})
\beta_0^{(n)}\beta_0^{(n+1)} .
\label{eqng0a1} 
\end{equation}
Thus, one may solve for the generating function $g_0 (a,b)$
from the kernelized equations in \Ref{eqnkernel1} and
\Ref{eqnkernel2}:
\begin{equation}
g_0 (a,b) = \frac{ab - tK\left(a^2g_0(a,0)+b^2g_0(b,0)\right)}{
1-KL}
\label{eqng0ab} 
\end{equation}
and the radius of convergence is given by the dominant root
of the quartic in equation \Ref{eqnquartic}, provided that
$g_0 (a,0)$ does not have compensating singularities at
the same points.

\subsection{More on the roots of $t^2(a^2+b^2)^2 - ab$.}

One may check that 
\begin{equation}
\beta_0 (a) = [2\,t^2a^3] \sum_{n=0}^{\infty} 
\left( {{4n+1}\atop{n}}\right) 
\frac{(at)^{4n}}{3n+2}
\label{eqnbeta01} 
\end{equation}
as given in equation \Ref{eqnbeta0}.  The radius
of convergence of this series is $|at|^4 = 3^3/4^4$.

Closer examination also shows that $\beta_0 (a)$ counts 
directed paths with first step in the SE-direction, 
above the line $Y=-X/2$ and with last vertex in the 
line $Y = -X/2$. The root $\beta_1$ is the inverse
function of $\beta_0$:  Direct calculation shows that
$\beta_0 \circ \beta_1 (a)= \beta_1 \circ \beta_0 (a)= a$.

The roots $\beta_+$ and $\beta_-$ are not independent,
but one may check that
\begin{equation}
\hspace{-2cm}
\beta_+ \left(\frac{ix^3}{8} \right)
= - \beta_- \left(\frac{-ix^3}{8} \right)
= - \frac{ix}{2t^{2/3}} + \frac{it^{2/3}x^5}{48} 
+ \frac{it^2x^9}{1536} + \frac{7it^{10/3}x^{13}}{165888} 
+ \ldots
\end{equation}

Solving for the roots of the quartic is equivalent to solving the
non-linear system
\begin{eqnarray}
r^2 & = & (a^2+b^2), 
\label{eqnr2} \\
t^2r^4 & = & ab
\label{eqnt2r4} 
\end{eqnarray}
for $a$ and $b$.  There are four solutions, two given by the pairs
$(a_1(r),b_1(r))$ and $(-a_1(r),-b_1(r))$, where
\begin{eqnarray}
a_1(r) & = & \frac{\sqrt{2}r^3t^2}{\sqrt{1+\sqrt{1-4t^4r^4}}}
= \frac{r\sqrt{1-\sqrt{1-4t^4r^4}}}{\sqrt{2}}; \\
b_1(r) & = & \frac{r\sqrt{1+\sqrt{1-4t^4r^4}}}{\sqrt{2}}
= \frac{\sqrt{2}r^3t^2}{\sqrt{1-\sqrt{1-4t^4r^4}}},
\end{eqnarray} 
and two more given by the pairs
$(b_1(r),a_1(r))$ and $(-b_1(r),-a_1(r))$ (where we interchanged
$a_1$ and $b_1$). An expression for $r^2$ is given below in
equation \Ref{eqnra2}.  One may check as well that
\begin{eqnarray}
a_1 (r) & = & t^2r^3\left( 1 + t^4r^4 \sum_{n=0}^\infty
\left( {{4n+3}\atop{2n}} \right) \frac{(tr)^{4n}}{(2n+1)4^{n+1/2}}
\right) 
\label{eqna1r}  \\
b_1 (r) & = & r\left(1 - t^4r^4 \sum_{n=0}^\infty
\left( {{4n+1}\atop{2n}} \right) \frac{(tr)^{4n}}{(2n+2)4^n}\right)
\label{eqnb1r} 
\end{eqnarray}

The roots of the quartic may be found by inverting $a_1$ to
obtain $r_a(a)$ so that $r_a \circ a_1 = a_1 \circ r_a$ is the
identity map.  Then $\beta_0 (a) = b_1\circ r_a$.  Inverting
$b_1$ to obtain $r_b (b)$ gives a second root by the composition
$\beta_1 = a_1 \circ r_b$.  In particular, this means for
example that $\beta_0 \circ \beta_1 = b_1\circ (r_a \circ a_1)
\circ r_b = b_1 \circ r_b = \hbox{identity}$ since 
$a_1^{-1} = r_a$ and $b_1^{-1} = r_b$.  This proves the observation
above that $\beta_0 \circ \beta_1 = \beta_1 \circ \beta_0$ is the
identity map.  In other words, the composition of two roots
of the quartic is the identity.

The other two roots of the quartic is given by the compositions
$b_1 \circ r_b$ and $a_1 \circ r_a$.  Unfortunately, while explicit
expressions for $r_a$ and $r_b$ can be obtained, they are rather
lengthy; both $r_a^2$ and $r_b^2$ are roots of the quartic
$t^4x^4 - c^2\,x+c^4$ where $c=a$ for $r_a$ and $c=b$ for $r_b$.
This may be examined by iteration to determine the first few
terms in $r^2_a$.  Comparison of the results to the online
encyclopedia of integers \cite{OL07} shows that
\begin{equation}
r_a^2 (a) = a^2 \sum_{n=0}^\infty 
\left( {{4n}\atop{n}} \right)
\frac{(at)^{4n}}{3n+1} .
\label{eqnra2} 
\end{equation}
The second root of $t^4x^4 - c^2\,x + c^4$ proposes the "unphysical"
series starting with $(a/t^2)^{2/3} + O(a^2)$ for $r_a^2 (a)$.
The series expression for $r_a^2 (a)$ may finally be substituted 
in $b_1 (r)$ to obtain an expression for the root $\beta_0(a)$
of the quartic:
\begin{equation}
\beta_0 (a) = \frac{r_a(a)\sqrt{1+\sqrt{1-4t^4r_a^4(a)}}}{\sqrt{2}}.
\end{equation}
Remarkably, this evaluates to equation \Ref{eqnbeta01} and
compositions of this with itself will eventually lead
to the expression for $g_0 (a,0)$ in equation \Ref{eqng0a1}.

In addition, having determined $\beta_0 (a)$, one may
consider the composition of $a_1 (r)$ and $r_a (a)$, which is
\begin{equation}
\frac{r_a (a)\sqrt{1-\sqrt{1-4t^4r_a^4(a)}}}{\sqrt{2}} = a,
\end{equation}
and which must be the identity map.  In other words, by appealing
to equation \Ref{eqnt2r4} it follows that
\begin{equation}
a \cdot \beta_0 (a) = t^2 r_a(a)^4,
\label{eqnbeta0ra4} 
\end{equation}
and from equation \Ref{eqnbeta01} one concludes that the identity
\begin{equation}
\left[ \sum_{n=0}^\infty 
\left( {{4n}\atop{n}} \right)
\frac{(at)^{4n}}{3n+1} \right]^2 = 
2 \sum_{n=0}^{\infty} 
\left( {{4n+1}\atop{n}}\right) 
\frac{(at)^{4n}}{3n+2} 
\end{equation}
should be closely related to the combinatorial properties of
directed paths in the wedge $V_2$.


\section{Determining $C_0$.}

In this section we examine the generating function 
$g_0 (a,b)$ in equation \Ref{eqng0ab} with $a=b=1$.  Singularities
in this generating function arise from several possible sources.
In the first instance there are simple poles at the zeros of
the kernel $(1-KL)$ in the denominator. These are located at
$t=\pm 1/2$.  In order to determine the constant $C_0$,
we examine the residue of $g_0 (1,1)/t^{n+1}$ at $t=\pm 1/2$. 

Put $a=1$ and $t=1/2$ in $\beta_0(a)$.  Compositions of
$\beta_0(a)$ with itself at this point gives the following
values
\begin{eqnarray*}
\beta_0^{(0)} (1) & = & 1 \\
\beta_0^{(1)} (1) & = & 2.9559774252208477098\ldots\times 10^{-1} \\
\beta_0^{(2)} (1) & = & 6.4633625443847777820\ldots\times 10^{-3} \\
\beta_0^{(3)} (1) & = & 6.7501832073150278963\ldots\times 10^{-8} \\
\beta_0^{(4)} (1) & = & 7.6892979457392165146\ldots\times 10^{-23} \\
\beta_0^{(5)} (1) & = & 1.1365801752937162161\ldots\times 10^{-67}\\
\beta_0^{(6)} (1) & = & 3.6706268625677440729\ldots\times 10^{-202}
\end{eqnarray*}
where explicitly
$$
\beta_0 (1) = \frac{5-\sqrt{33}}{12}(19+3\sqrt{33})^{2/3} + 
  \frac{\sqrt{33}-1}{12}(19+3\sqrt{33})^{1/3} -
  \frac{1}{3} ,
$$
and the other terms are more complicated expressions involving
nested radicals which do not simplify to manageable expressions.

Generally, we observe that $\beta_0^{(n)} (1)
= \alpha_n$ at $t=1/2$, and one may check that
$\alpha_{n+1} \approx \alpha_n^3/4$.  For example,
$\alpha_3^3/4 = 7.689297941\ldots\times 10^{-23} \approx
\alpha_4$.  This observation follows from the expansion
$\beta_0 (a) = a^3t^2 + O(a^7t^6)$ so that compositions
of $\beta_0$ at $t=1/2$ and $a=1$ quickly converges to zero.
In other words, the recurrence $x_{n+1} = \beta_0(x_n)|_{t=1/2}$ 
is a fixed point iteration of order three.
This fast convergence allows the accurate numerical estimation 
of $\beta_0^{(n)}(1)$ at $t=1/2$.

At $a=b=1$ the expression of $g_0 (1,1)$ in terms of
$g_0 (1,0)$ and $g_0 (0,1)$ is given by,
\begin{equation}
g_0 (1,1) = \frac{1-2t^2(g_0(1,0)+g_0(0,1))}{1-4t^2}.
\label{eqng011} 
\end{equation}
where
$$g_0 (1,0) = g_0 (0,1) = 
\sum_{n=0}^\infty (-1)^n L(\beta_0^{(n)})
\beta_0^{(n)} \beta_0^{(n+1)},$$
and where $L(a) = a^2 + \beta_0^2$.
Numerical evaluation of the residue at $t=1/2$ using 
the calculated values of $\beta^{(n)}(1)$ above, gives 
the leading order behaviour of the number of paths of length 
$n$:
$$c_n = 2^{n-1} \times 0.6787405307981094574172327\ldots
 + \hbox{parity term} + \ldots$$
where the seven values of $\beta_0^{(n)}(1)$ listed above
produces $C_0$ accurately up to at least $O(10^{-401})$
or $400$ digits if each is calculated to at least this
accuracy by using equations \Ref{eqnbeta0ra4}
or \Ref{eqnbeta01} or by explicitly using the closed form
expression for $\beta_0 (a)$ and computing it to high
accuracy using a symbolic computations package such as
Maple.

To determine the parity effects, we determine the residue at the 
pole located at $t=-1/2$ by repeating the analysis above.  Put
$a=1$ and $t=-1/2$ in $\beta_0 (a)$.  Compositions of
$\beta_0 (a)$ with itself at this point gives the identical
values obtained above for $t=1/2$.  Thus, we conclude that
$$c_n = 2^{n-1}(1+(-1)^n)
\times (0.6787405307981094574172327\ldots)
  + \ldots$$
This is not an unexpected result since $g_0(1,1)$ should
only enumerate even length paths; we note that
$c_{2n+1} = 0$ in the generating function $g_0 (1,1)$.  This
result verifies the numerical estimate for $C_0$ from
the data in table \ref{table1} in equation \Ref{eqn7}.

Contributions to $c_n$ also arise from singularities
in the numerator in equation \Ref{eqng0ab}.  In particular,
there are branch points and possibly other singularities
in $g_0 (1,0)$ and $g_0 (0,1)$, and these are due to 
branch points in $\beta_0$.

The radius of convergence of $\beta_0$ can be determined
when $a=1$ by examining equations \Ref{eqnra2} and
\Ref{eqnbeta0ra4} with $a=1$.  In particular, $r_1(1)$
in equation \Ref{eqnra2} is convergent for all $|t|
\leq 3^{3/4}/4$; in fact, evaluation
shows directly that $r_1 (1)$ is convergent for
$|t| = 3^{3/4}/4$.  Since $3^{3/4}/4 > 1/2$, this proves that
the simple poles at $t=\pm 1/2$ are within the radius of
convergence of $\beta_0 (1)$.

Further examination of $\beta_0 (a)$ by Maple shows a complicated
expression of nested radicals which expicitly contains
factors of the form $\sqrt{81-768 a^4t^4}$.  This
shows that there are branch points at $at = 3^{3/4}\omega/4$
where $\omega$ is a fourth root of unity.  
There may be more branchpoints on the circle
$at = 3^{3/4}/4$, but we did not verify this, and this will
not play a crucial role in what follows.

Next, we consider the compositions of $\beta_0$ in the
alternating sum definition of $g_0 (1,0)$. Since
$r_1(t)$ is a positive term power series it follows
by the triangle inequality in equation \Ref{eqnra2} that
for (complex) $t$ such that $|t| \leq 3^{3/4}/4$,
\begin{equation}
|r^2_1(1)| \leq \sum_{n=0}^\infty
\left( {{4n}\atop{n}} \right)
\frac{(3^{3/4}/4)^{4n}}{3n+1} = \frac{4}{3}.
\end{equation}
Thus, by equation \Ref{eqnbeta0ra4}
\begin{equation}
|\beta_0 (1) | \leq \left[3^{3/4}/4\right]^2(16/9)
= \frac{1}{\sqrt{3}}, 
\qquad \hbox{if $|t| \leq 3^{3/4}/4$.}
\end{equation}
Since $\beta_0 (a)$ is a power series with positive coefficients
in both $a$ and $t$, $|\beta_0 (a)| \leq |\beta_0 (1)|$ for
any $|t| \leq 3^{3/4}/4$ and $|a|\leq 1$, and by the 
triangle inequality, $\beta_0 (a)$ is a maximum when 
$t=3^{3/4}/4$ for a fixed value of $a$.  Thus, for fixed
values of $|a|\leq 1$, $\beta_0 (a)$ is a maximum on the closed
disk $|t|\leq 3^{3/4}/4$ when $t=3^{3/4}/4$.

The above shows that for $t=3^{3/4}/4$, $|\beta_0 (a)| 
\leq 1/\sqrt{3}$ for $|a|\leq 1$.  Moreover, it
follows from equation \Ref{eqnbeta01} that
$|\beta_0 (a)| \leq |a|/\sqrt{3}$ for $|a| \leq 1$ when
$t=3^{3/4}/4$.  In other words,
\begin{equation}
|\beta_0 (\beta_0 (1)) |
\leq |3^{3/4}/4\sqrt{3}(\sqrt{3}\cdot3^{3/4}/4)|
= \frac{1}{\sqrt{3}^2}.
\end{equation} 
It follows inductively that
\begin{equation}
\left| \beta_0 ^{(n)} (1) \right| \leq \frac{1}{\sqrt{3}^n}.
\label{eqnbetasq3} 
\end{equation}
Since the branch points in $\beta_0 ^{(n)}(1)$ will occur at
values of $t$ that $|\beta_0^{(n-1)} (1)| = 3^{3/4}/4$, and
$|\beta_0^{(n-1)}(1)| \leq 1/\sqrt{3}^{n-1} < 3^{3/4}/4$ at 
$a=1$ and for $n>2$, this implies that the branch points
in $\beta^{(n)}(1)$ for $n>1$ lies outside the circle with radius
$|t| = 3^{3/4}/4$, and contributions of these branch points
to $c_n$ are dominated by the contributions to $c_n$ which
is a result of the branch points in $\beta_0(1)$ itself.

In particular, since the radius of convergence of $\beta_0 (a)$
is determined by $|at| = 3^{3/4}/4$ and $|\beta_0^{(n-1)}(1)|
\leq 1/\sqrt{3}^{n-1}$, it follows that the radius of convergence
of $\beta_0^{(n)}(1)$ in the $t$-plane is on or outside the
circle with radius $|t| = 3^{3/4}\sqrt{3}^{n-1}/4$, for $n>2$.
Thus, the generating function $g_0 (a,0)$ has infinitely many
singularities in the $t$-plane for $a=1$, and by 
remark 1 following proposition 9 in reference \cite{BMP03},
$g_0 (1,0)$ cannot be holonomic (or D-finite).  Thus, 
$g_0 (1,1)$, the generating function of even length 
directed paths in the wedge formed by $Y=\pm X/2$, is 
not holonomic.

The bound in equation \Ref{eqnbetasq3} also
proves that $g_0 (1,0)$ is an absolutely convergent series
on the open disk with radius $|t| = 3\sqrt{3}$ which includes the
simple poles of $g_0(1,1)$ at $t=\pm 1$ and the branch points
on the circle $|t| = 3^{3/4}/4$ in its interior.
In other words, the asymptotic behaviour of $c_n$ is given by
\begin{eqnarray}
c_n & = & 2^{n-1}(1+(-1)^n)\times \
(0.6787405307981094574172327\ldots)\nonumber \\
& & + \hbox{corrections.}
\label{eqncnres} 
\end{eqnarray}
The corrections are due to the branch points in 
$\beta_0(1)$ and they grow at the exponential rate
$(4/3^{3/4})^{n+o(n)}$. Since $4/3^{3/4} \approx 1.75 < 2$, 
the effects of the correction terms will dissappear fast 
with increasing $n$, and $c_n/2^n$ will approach 
$0.6787405307981094574172327\ldots$
at an exponential rate with increasing $n$.

\section{Narrower Wedges}

It is possible to consider this problem in the narrower wedge
$V_p$ where $a$ and $b$ measure vertical distance to the
functions $Y = \pm \lceil X/p \rceil$ and where the path
starts at the vertex with coordinates $(p,0)$.  In this
case the paths are counted by a generating function
$G_p (a,b) = \sum_{i=0}^{p-1} g_i (a,b)$ where $g_i (a,b)$
generates paths of length $i \mod p$ and satisfies the
set of coupled recurrance equations with a derivative term:
\begin{eqnarray*}
g_0 = & &
ab + t(a/b+b/a)\,g_{p-1}, \\
g_1 = & & 
t(a^2+b^2)\,g_0 - ta^2g_0(a,0) - tb^2g_0(0,b), \\
g_2 = & & 
t(a/b+b/a)\,g_1 
- ta \left[\frac{\partial g_{1}}{\partial b}\right]_{b=0}
- tb \left[\frac{\partial g_{1}}{\partial a}\right]_{a=0} \\
g_3 = & & 
t(a/b+b/a)\,g_2 
- ta \left[\frac{\partial g_2}{\partial b}\right]_{b=0}
- tb \left[\frac{\partial g_2}{\partial a}\right]_{a=0} \\
\ldots = & & \ldots \\
g_{p-1} = & & 
t(a/b+b/a)\,g_{p-2}
- ta \left[\frac{\partial g_{p-2}}{\partial b}\right]_{b=0}
- tb \left[\frac{\partial g_{p-2}}{\partial a}\right]_{a=0} 
\end{eqnarray*}
Putting $p=2$ in the above recovers the recurrences in 
equation \Ref{eqnrecur2}, if the fact that $g_1 (a,0) 
= g_1(0,b) = 0$ is used.

Putting $a=b=1$, writing $G = \sum_{i=0}^{p-1} g_i$ 
and defining the boundary terms $\Delta g_0 (a,0) = 0$, 
$\Delta g_1(a,0) = g_1(a,0)$ and $\Delta g_j (a,0) =  
\left[\frac{\partial g_j}{\partial b}\right]_{b=0}$ for 
$j=2,3,\ldots,p-1$ and similarly for $\Delta g_j (0,b)$,
show that by summing the above recurrances one obtains
\begin{equation}
G = 1 + 2t\, G - \sum_{i=0}^{p-1} \left(
\Delta g_i (1,0) + \Delta g_i (0,1) \right) . 
\label{eqn38} 
\end{equation}
The boundary terms $\sum_{i=0}^{p-1} \left(
\Delta g_i (1,0) + \Delta g_i (0,1) \right)$ represent the generating
function of paths starting at the point $(p,0)$ and ending
within one step from the lines $Y=\pm X/p$. Such paths of
length $2pn$ must give almost (within a constant) $(p+1)n$
North East steps.  The number of these paths grows at the exponential
rate $\lambda = ((2p)^{2p}/(p-1)^{p-1}(p+1)^{p+1})^n$.  Hence the
generating function of these paths is convergent inside
the circle of radius $1/\lambda^{1/2p}$.  One can show that
this is strictly greater than $1/2$, hence all the boundary
generating functions $\Delta g_i (1,0)$ and $\Delta g_i (0,1)$
of paths ending near the boundary of the wedge
are convergent on the disk with radius strictly greater than $1/2$.

Thus, rearranging equation \Ref{eqn38} to get
\begin{equation}
G = \frac{1-\sum_{i=0}^{p-1} \left(
\Delta g_i (1,0) + \Delta g_i (0,1) \right)}{1-2t}
\end{equation}
the dominant singularity in $G$ is at $t=1/2$ and is a simple
pole.  The subdominant singularities give terms growing 
to exponential order $\lambda^{n+o(n)}$ where
\begin{equation}
\lambda = \left((2p/(p-1)^{(p-1)/2p}(p+1)^{(p+1)/2p})^n\right)
\end{equation}
 In other words,
\begin{equation}
c_n^{(p)} = A_p 2^n + O\left(\lambda^n\right).
\end{equation}
Putting $p=2$ produces the leading order term and the order of
the next term obtained in equation \Ref{eqncnres}.

\section{Conclusions}

The main results in this paper are given by equations
\Ref{eqng0ab} and \Ref{eqncnres}.  We have shown that
the model of fully directed paths in the wedge $V_2$
in figure \ref{fig3} can be solved using the iterated kernel
method.  This model is the second in a sequence of models
in the wedge $V_p$, for $p\in\{1,2,3,\ldots\}$ and recurrences
for $p>2$ is given in section 2.4 (when $p=1$ the model is trivial, and $c_n = 2^n$). We proved that the number
of fully directed path of length $n$ in $V_2$ increases
exponentially at the rate $C_0 2^n$, plus subdominant terms 
which are also exponentials.  In addition, we showed that $C_0$ (the
coefficient of the leading order term) can be determined
to high accuracy.  

In the more general wedge $V_p$, it may be shown, using
the techniques in reference \cite{HW85}, that the number of
paths of length $n$ is $c_n^{(p)} = C_0^{(p)} 2^n +
\hbox{lesser terms}$.  In this event $C_0^{(p)}$ is known
for $p=1$ ($C_0^{(1)}=1$) and to high accuracy for $p=2$
($C_0^{(2)} = 0.678740\ldots$).  No estimates exist for
other values of $p$.

The ``physical" root $\beta_0 (a)$ of the quartic kernel 
(see equation \Ref{eqnbeta01}) counts directed paths from
$(2,0)$ (see figure \ref{fig3}) above the line $Y=-X/2$ and
with last vertex in the line $Y=-X/2$.  By reversing the
horizontal direction, this is also equal to the number of
paths from the origin above the line $Y=X/2$ and with last
vertex in the line $Y=X/2$.  The generating function in equation
\Ref{eqng0a1} is an alternating series of products of
compositions of $\beta_0 (a)$, suggesting that that
$g_0(a,0)$ is ``constructed" by an inclusion-exclusion
process of directed paths above the lines $Y=-X/2$ and $Y=X/2$.

These observations suggest 
that a combinatorial explanation for the
generating function $g_0 (a,b)$ in equation \Ref{eqng0ab}
might be possible in terms of paths ``bouncing off"
the boundaries of the wedge $V_2$.  We have not been able
to find such an explanation.  We leave this question open, but do 
observe that an explicit explanation in this case might
give clues about the nature of the generating function
of models in more general wedges, including the models
of directed paths in the wedges $V_p$ for $p>2$ formed
by the lines $Y=\pm X/p$, as well as in the half-wedges
formed by the $X$-axis and the lines $Y=X/p$. 
Such models in half-wedges will be more difficult to
solve, since the symmetric nature of $V_2$ plays
a key role in writing down equation \Ref{eqng0ab}
and then determining $C_0$ by examining the singularities
in equation \Ref{eqng011}.

Finally, our models in the wedge $V_p$ map to the 
well-known problem of random walks with given step-sets
in the quarter plane (the first quadrant).  These have been
studied for example in references \cite{M02,BMP03,MR07}. 
The models in this paper correspond to models with starting 
vertex $(p-1,p-1)$ and step set $\{(1-p,p+1),(p+1,1-p)\}$.  For 
$p=2$, this produces the generalised knight's walk model
with step set $(-1,3)$ and $(3,-1)$ \cite{JvRRP07}.  The knight's
walk problem itself (with step set $\{(-1,2),(2,-1\})$ and
starting vertex $(1,1)$ was studied in reference
\cite{BMP03}, where it was shown that its generating function
is not holonomic (or $D$-finite).

\vskip 8mm

\noindent{\bf Acknowledgments:} EJJvR and AR are supported by
Discovery Grants from NSERC (Canada).

\vspace*{0.5cm}

\noindent{\bf{Bibliography}}

\vspace*{0.3cm}

\begin{center}

\end{center}

\end{document}